\documentclass[conference]{IEEEtran}
\usepackage{setspace}
\usepackage{clrscode}
\usepackage{amsmath}
\usepackage{amssymb}
\usepackage[dvips]{graphicx}
\usepackage{epsfig}
\usepackage{hyperref}
\usepackage{bm}
\usepackage{subfigure}
\usepackage{indentfirst}
\usepackage{float}
\usepackage{stfloats}
\usepackage{multirow}

\newcommand{\y}{\mathbf{y}}
\newcommand{\h}{\mathbf{h}}
\newcommand{\B}{\mathbf{B}}
\newcommand{\z}{\mathbf{z}}
\newcommand{\w}{\mathbf{w}}
\newcommand{\U}{\mathbf{U}}
\newcommand{\dd}{\mathbf{s}}
\newcommand{\bb}{\mathbf{b}}
\newcommand{\figsize}{0.4}
\newcommand{\PP}{\mathbf P}
\newcommand{\p}{\mathbf p}

\newcommand{\HH}{\mathbf{H}}

\newcommand{\Q}{\mathbf{Q}}
\newcommand{\C}{\mathbf{C}}
\newcommand{\V}{\mathbf{V}}

\newcommand{\tout}{\text{\footnotesize{out}}}

\newcommand{\ssnr}{\text{\scriptsize{SINR}}}

\newtheorem{Lem}{Theorem}

\newtheorem{Corr}{Corollary}

\begin{document}
\title{Outage Analysis of Heterogeneous mmWave Cellular Networks Employing JSDM}
    \author{
\IEEEauthorblockN{Jun Chen and Deli Qiao}
\IEEEauthorblockA{School of Information Science and Technology, \\East China Normal University, Shanghai, China\\
Email: 51161214011@stu.ecnu.edu.cn, dlqiao@ce.ecnu.edu.cn}}
\maketitle

\begin{abstract}
In this paper, the outage performance of a two-tier heterogeneous mmWave cellular networks employing joint spatial division and multiplexing (JSDM) is investigated. It is assumed that macro base stations (BSs) equipped with a large number of antennas and pico BSs equipped with single antenna serve users simultaneously. The two-tier BSs and users are distributed according to independent Poisson point processes (PPPs). Theoretical analysis of the signal-to-interference-plus-noise ratio (SINR) outage probability of the typical user is first provided. The all SINR outage probability of the two tiers is then derived. Simulation results in accordance with theoretical analysis demonstrating the performance improvement of two-tier networks compare with the single-tier network are provided. By using the noise-limited assumption for mmWave networks, a simpler expression to analyze the outage performance is obtained.
\end{abstract}

\section{Introduction}
 Massive multiple input multiple output (MIMO) \cite{tom-noncooperative}, \cite{massiveMIMO}, millimeter wave (mmWave) \cite{tsrapp}, \cite{2},  and ultra-dense networks  \cite{1}, \cite{udn} have been proposed to address the exponentially increased demands in mobile data traffic. MmWave communication systems can offer an order of magnitude increase in achievable rate compared with current cellular systems and play an important role in future cellular networks \cite{1}.

Unfortunately, it is highly possible that some users may fall in outage in mmWave systems, because the transmission range is generally limited due to the high free-space path loss and poor penetration in high frequency \cite{tsrapp}. Hence there are some studies about the coverage analysis of mmWave systems recently (see e.g. \cite{stommWave}-\cite{wcsp} and references therein). For instance, an analytical framework which computes coverage probabilities and rate of mmWave cellular networks has been proposed in \cite{stommWave}, where path-loss and blockage models based on empirical data for mmWave propagation have been taken into account. Relying on the noise-limited assumption for modeling mmWave cellular systems, simple and closed-form formulas for computing the coverage probability and the average rate have been obtained. The authors have proposed a stochastic geometry framework for analyzing the coverage and rate of mmWave systems assuming pencil beams for the users in \cite{coverage}. They have shown that there is an optimal relative base station (BS) density for the signal-to-interference-plus-noise ratio (SINR) and rate performance beyond which the performance does not improve in a dense mmWave network. In \cite{gursoycoverage}, the authors have made a further step and investigated the coverage in heterogeneous mmWave networks with homogeneous Poisson point process (PPP) models of the BS and user distributions, where beamforming with pencil beam at the BSs is assumed. It has been shown that biasing towards the small cells in user association can improve both the coverage probability and the rate. Also, the authors have assumed homogeneous PPP model for the macro cells while Poisson hole process (PHP) model for small cells and studied the coverage of the proposed non-uniform mmWave heterogeneous cellular network in \cite{wcsp}.

In this paper, we investigate the outage performance of a two-tier heterogenous mmWave cellular networks employing joint spatial division and multiplexing (JSDM). For mmWave systems employing massive MIMO, a two-stage precoding scheme, JSDM has been proposed in \cite{jsdmmmwave}. The idea of JSDM is to make use of the channel covariance information to reduce the channel estimation overhead and mitigate the interference for users in different groups partitioned according to channel covariance subspaces \cite{jsdm}. In this paper, the two-tier BSs and users are assumed to be distributed according to independent PPPs. We first exploit zero forcing (ZF) at the macro BS as the second precoding matrix to cancel the interference from the user in the same group. Then we obtain the user SINR outage probability and cell SINR outage probability.

The paper is organized as follows. The system model and the preliminaries on the user association and JSDM are briefly introduced in Section II. Section III discusses the SINR outage analysis of the two-tier mmWave systems in detail. Numerical results are provided in Section IV. Finally, Section V concludes this paper.

\section{System Model}
In this section, a two-tier heterogeneous downlink  mmWave cellular network is modeled. The BSs in the $k^{th}$ tier are distributed according to a homogeneous PPP $\Phi_k$ of density $\lambda_k$ on a region $A_k$ for $k=1,2$. The BSs in the same tier $k$ transmit with the same power $P_k$ for $k=1,2$ and all BSs in the two tiers transmit in a mmWave frequency band. The users are also distributed according to a homogeneous PPP $\Phi_u$ of $\lambda_u$. The BSs in different tiers and the users are all independent. In our model, a typical user is assumed to be located at the origin.  The macro BS equipped with $M$ antennas with uniform linear array (ULA) serves the users and pico BSs equipped with one antenna.

We adopt a line-of-sight (LOS) ball model according to \cite{coverage}. In this model, the path loss are $r_{\ell}^{-\alpha_s}$ and $r_{\jmath}^{-\alpha_s}$ in tier 1 and tier 2, respectively, $s\in (L,N)$ denotes the typical user can be either a LOS or non-line-sight (NLOS) link. The probabilities of the LOS link in tier 1 and tier 2 are given by
\begin{align}
p_L^1(x)&=\left\{
                \begin{aligned}
                    &1 &\quad \mathrm{if}\quad x<R_L\\
                    &0  &otherwise
                \end{aligned},
                \right.\\
p_L^2(x)&=\left\{
                \begin{aligned}
                    &1 &\quad \mathrm{if}\quad x<r_L\\
                    &0  &otherwise
                \end{aligned}.
                \right.
\end{align}

\subsection{Cell Association}
 According to \cite{StochasticGeometry}, a Poisson distributed number $n_k$ with mean $\Lambda_k=\lambda_k\mid A_k\mid$ and $n_k$ points is distributed uniformly at random in $A_k$.
Let $r_\ell$ denotes the distance between the macro BS and the typical user , $\ell=1,..., n_1$, and $r_\jmath$ denotes the distance between pico BS and the typical user, $\jmath=1,..., n_2$. The probability density functions (pdf) of the distances are $f_1(r_\ell)=\frac{2r_\ell}{R^2}, \ell=1,..., n_1$ and $f_2(r_\jmath)=\frac{2r_\jmath}{r^2}, \jmath=1,..., n_2$. $R$ and $r$ are the radiuses of tier 1 and tier 2, respectively.

According to \cite{UltraDense}, the probability density functions of the smallest distances $r_{m}$ and $r_{s}$ are given by
\begin{align}
f_m(r_m)&=2\lambda_m\pi r_me^{-\lambda_m\pi r_m^2},\\
f_s(r_s)&=2\lambda_s\pi r_se^{-\lambda_s\pi r_s^2}.
\end{align}
Assuming equal power allocation among the data streams, a typical user is associated with a pico BS if
\begin{align}
P_s\kappa^2r_{s}^{-\alpha_s}\geq P_m\kappa^2r_{m}^{-\alpha_s},
\end{align}
where $P_m=\frac{P_1}{S}$ and $S$ denotes the number of users served by the macro BS, $P_s=P_2$ and $\kappa^2=(\frac{\lambda_c}{4\pi})^2$.

\subsection{Joint Spatial Division and Multiplexing with Per-Group Processing (JSDM-PGP)}

We consider a one-ring scattering model for the channel between the users and the macro BSs and users and the pico BSs, respectively. Taking into account the small scale fading only, the channel covariance matrix for a user in group $g$ with angle-of-arrial (AOA) $\theta_g$ and angular spread (AS) $\Delta_g$ is given by \cite{jsdmmmwave}
\begin{align}
[\mathbf{R}_g]_{m,p}=\frac{1}{2\Delta_g}\int_{-\Delta_g+\theta_g}^{\Delta_g+\theta_g}e^{-j2\pi D(m-p)\sin(t)}\, dt,
\end{align}
where $D$ denotes the distance between the adjacent antenna elements of the macro BS in terms of carrier wavelength. Assume that the eigenvalue decomposition of $\mathbf{R}_g$ is given by $\mathbf{R}_g=\U_g\boldsymbol{\Lambda}_g\U_g^H$, where $\mathbf{U}_g$ is a tall unitary matrix of dimensions $M\times r_g$, $\boldsymbol{\Lambda}_g$ is a $r_g\times r_g$ diagonal semi-positive definite matrix, and $r_g$ denotes the rank of $\mathbf{R}_g$. Thus, the small-scale fading channel of user $k$ in group $g$ can be written without loss of generality as
\begin{align}\label{eq:channel}
\h_{gk}=\U_{g}\boldsymbol{\Lambda}_g^{1/2}\w_{gk},
\end{align}
where $\w_{gk}\sim\mathcal{CN}(\boldsymbol{0},\boldsymbol{I}_{r_g})$ is an i.i.d. Gaussian random vector.

We assume that JSDM-PGP, a two-stage transmission scheme with dimension reduced channel state information, is employed at the macro BS. In this scheme, $K$ users are divided into $G$ groups with $K_{g}$ users each group and $K=\sum_{g=1}^GK_{g}$, and the received signal at the users in group $g$ is given by
\begin{align}
\y_g=\HH_g^H\B_g\PP_g\dd_g+\sum_{g'\neq g}\HH_g^H\B_{g'}\PP_{g'}\dd_{g'}+\z_g,
\end{align}
where $\B_g=[\bb_{g1},\ldots,\bb_{gB_g}]$ is the first-stage precoding matrix of dimension $M\times B_g$ to reduce the dimension of the channel and null the inter-group interference, and $\PP_g$ is the second stage precoding matrix of dimension $B_g\times S_g$. $S_g$ denotes the number of data streams in group $g$ and $S_g\le B_g\le r_g$. Denote $S=\sum_{g}S_g$ as the total number of data streams. $\HH_g=[\h_{g1},\ldots,\h_{gK_g}]$ is composed the instantaneous channel state information of the users in group $g$. $\dd_g \in \mathbb{C}^{K_g\times1}$ is the transmitted signal for the users in group $g$ and $\z_g \in \mathbb{C}^{K_g\times1}$ is additive white Gaussian noise at the users with i.i.d. entries of zero mean and unit variance.

Generally, $\B_g$ has been designed based on the long-term channel statistics to null the inter-group interference, i.e., $\HH_g^H\B_{g'}\approx \mathbf{0}$, for all $g_{'}\neq g$ \cite{jsdmmmwave}. $\PP_g$ is decided by the equivalent channel $\HH_g^H\B_g$ seen by the users in group $g$.

\section{SINR Outage Probability}
In this section, we investigate the SINR outage probability for the two-tier heterogenous mmWave cellular networks employing JSDM. We first derive the ZF matrix as the second precoding to obtain the expression of SINR and then find the user SINR outage probability and the all SINR outage probability.
\subsection{Signal-to-Interference-Plus-Noise Ratio (SINR)}

In \cite{jsdmmmwave}, the effective channel is $\underline{\h}_{m,gk}=\B_{m,g}^H\h_{m,gk}$, e.g., $\underline{\h}_{m,gk}=\B_{m,g}^H\U_{m,g}\boldsymbol{\Lambda}_{m,g}^{1/2}\w_{m,gk}$. After some calculations, we obtain $\C_{m,g}=\mathrm{diag(diag}(\V_{m,g}\V_{m,g}^H))^{1/2}$ and $\V_{m,g}=\B_{m,g}^H\U_{m,g}\boldsymbol{\Lambda}_{m,g}^{1/2}$ to let $\C_{m,g}^{-1}\B_{m,g}^H\h_{m,gk}\sim \mathcal{CN}(0,\mathbf{I})$. In our work, the effective channel is defined as
\begin{align}
 \overline{\h}_{m,gk}=\frac{\C_{m,g}^{-1}\B_{m,g}^H}{||\C_{m,g}^{-1}\B_{m,g}^H||}\h_{m,gk}.
\end{align}

Taking into account the path loss effect, the received signal of the typical user in group $g$ served by the macro BS $m$ can be expressed as
\begin{align}
y_{mk}&=\underbrace{\kappa r_{m}^{-\alpha_s/2}\overline{\h}_{m,gk}^{H}\p_{m,gk}s_{m,gk}}_{\text{useful signal}}\nonumber\\
&+\sum_{k'\neq k}\underbrace{\kappa r_{m}^{-\alpha_s/2}\overline{\h}_{m,gk}^{H}\p_{m,gk'}s_{m,gk'}}_{\text{inter group interference}}\nonumber\\
&+\sum_{g'\neq g}\underbrace{\kappa r_{m}^{-\alpha_s/2}\h_{m,gk}^{H}\B_{m,g'}\PP_{m,g'}\dd_{m,g'}}_{\text{intral group interference}}\nonumber+
\end{align}
\begin{align}
&\underbrace{\sum_{\ell\ne m}\kappa r_{\ell}^{-\alpha_s/2}\overline{\h}_{\ell}^{H}\p_{\ell}s_{\ell}}_{\text{intra BSs interference in same tier}}
+\underbrace{\sum_{\jmath}\kappa r_{\jmath}^{-\alpha_s/2}h_{\jmath}s_{\jmath}}_{\text{intra BSs interference in intra tier}}
+\underbrace{z_{k}}_{\text{noise}},
\end{align}
where $y_{mk}$ denotes the received signal of the typical user $k$ from the serving macro BS, $\overline{\h}_{\ell}^{H}$ denotes the effective channel from intra BSs in the same tier, $h_{\jmath}$ denotes the channel form intra BSs in intra tier and $h_{\jmath}\sim \mathcal{CN}(0,1)$. $s_{\ell}, \ell=1,..., n_1$ are the sent signals from the intra macro BSs in the same tier for the typical user and $s_{\jmath}, \jmath=1,..., n_2$ are the sent signals from the intra BSs interference in intra tier for the typical user. According to \cite{interference}, the zero forcing matrix is defined as
\begin{align}
\p_{m,gk}=\frac{\Q_{m,gk}\Q_{m,gk}^{H}\overline{\h}_{m,gk}}{||\Q_{m,gk}\Q_{m,gk}^{H}\overline{\h}_{m,gk}||},
\end{align}
where $\Q_{m,gk}=\mathrm{null}(\HH_{m,gk})$, $\HH_{m,gk}=[\overline{\h}_{m,g1},...,\overline{\h}_{m,gi},...,\overline{\h}_{m,gK_g}],i\neq k$ and $\mathrm{\PP_{m,g'}=[\p_{m,g'1}...\p_{m,g'K_{g'}}]}$.

The received signal of the typical user in group $g$ served by the pico BS can be expressed as
\begin{align}
y_{sk}&=\underbrace{\kappa r_{s}^{-\alpha_s/2}h_{sk}^{H}s_{sk}}_{\text{useful signal}}
+\underbrace{\sum_{\ell}\kappa r_{\ell}^{-\alpha_s/2}\overline{\h}_{\ell}^{H}\p_{\ell}s_{\ell}}_{\text{intra BSs interference in intral tier}}\nonumber\\
&+\underbrace{\sum_{\jmath\neq s}\kappa r_{\jmath}^{-\alpha_s/2}h_{\jmath}^{H}s_{\jmath}}_{\text{intra BSs interference in same tier}}
+\underbrace{z_{k}}_{\text{noise}}.
\end{align}

Due to the two-stage precoding scheme, $|\overline{\h}_{m,gk}^{H}\p_{m,gk'}|^2=0$ and $||\h_{m,gk}^{H}\B_{m,g'}\PP_{m,g'}||^2=0$.
The SINR of a typical user served by the macro BS or the pico BS can be expressed as
\begin{small}
\begin{align}
&\ssnr_{mk}=\frac{P_m\kappa^2r_{m}^{-\alpha_s}\left|\overline{\h}_{m,gk}^{H}\p_{m,gk}\right|^2}{N_0+I_1+I_2},\\
&\ssnr_{sk}=\frac{P_s\kappa^2r_{s}^{-\alpha_s}\left|h_{s,gk}\right|^2}{N_0+I_3+I_4},
\end{align}
\end{small}
where $N_0$ is the noise power. $I_1, I_2$ and $I_3, I_4$ denote the interference from different BSs and they are defined as
\begin{align}
I_1&=\sum_{\ell\ne m}P_m\kappa^2 r_{\ell}^{-\alpha_s}\left|\overline{\h}_{\ell}^{H}\p_{\ell}\right|^2,\\
I_2&=\sum_{\jmath}P_s\kappa^2r_{\jmath}^{-\alpha_s}\left|h_{\jmath}\right|^2,\\
I_3&=\sum_{\ell}P_m\kappa^2 r_{\ell}^{-\alpha_s}\left|\overline{\h}_{\ell}^{H}\p_{\ell}\right|^2,\\
I_4&=\sum_{\jmath\neq s}P_s\kappa^2r_{\jmath}^{-\alpha_s}\left|h_{\jmath}\right|^2.
\end{align}

\subsection{User SINR Outage Probability}
First, we can show the following result regarding the outage probability of the typical user served by the macro BS.

\begin{Lem}\label{prop:SINRoutage}
The outage probability of the user served by the macro BS and pico BS are given by
\begin{align}
P_{out,s}^{m}(T) &= 1-e^{-\xi_m r_m^{\alpha_s}T}\mathcal{L}_{I_1}\mathcal{L}_{I_2},\\
P_{out,s}^{s}(T) &=  1-e^{-\xi_s r_s^{\alpha_s}T}\mathcal{L}_{I_3}\mathcal{L}_{I_4},
\end{align}
where
\begin{small}
\begin{align}
\xi_m&=\frac{N_0||\C_{m,g}^{-1}\B_{m,g}^H||^2}{P_m\kappa^2}, \quad \xi_s=\frac{N_0}{P_s\kappa^2},
\end{align}
\end{small}
\begin{small}
\begin{align}
\mathcal{L}_{I_1}&=\mathrm{exp}\Bigg(-\int_{r_m}^{\Re_m}2\pi\lambda_m\alpha_sx^{\alpha_s}\Big(1-\frac{1}{1+\xi_m r_m^{\alpha_s}TP_k\kappa^2x^{-\alpha_s}}\Big)\,dx\Bigg),\\
\mathcal{L}_{I_2}&=\mathrm{exp}\Bigg(-\int_{r_m}^{\Re_m}2\pi\lambda_s\alpha_sx^{\alpha_s}\Big(1-\frac{1}{1+\xi_m r_m^{\alpha_s}TP_s\kappa^2x^{-\alpha_s}}\Big)\,dx\Bigg),\\
\mathcal{L}_{I_3}&=\mathrm{exp}\Bigg(-\int_{r_s}^{\Re_s}2\pi\lambda_m\alpha_sx^{\alpha_s}\Big(1-\frac{1}{1+\xi_s r_s^{\alpha_s}TP_k\kappa^2x^{-\alpha_s}}\Big)\,dx\Bigg),\\
\mathcal{L}_{I_4}&=\mathrm{exp}\Bigg(-\int_{r_s}^{\Re_s}2\pi\lambda_s\alpha_sx^{\alpha_s}\Big(1-\frac{1}{1+\xi_s r_s^{\alpha_s}TP_s\kappa^2x^{-\alpha_s}}\Big)\,dx\Bigg),
\end{align}
\end{small}
where $\Re_m$ denotes the upper bound of $r_m$ according to $L^{m}_i$ and $\Re_s$ denotes the upper bound of $r_s$ according to $L^{s}_i$.
\end{Lem}
\emph{Proof:} See Appendix \ref{app:SINRoutage} for details.\hfill$\square$

\subsection{The All SINR Outage Probability}
According to PPP, there is a case that no BSs in the two-tier heterogeneous downlink  mmWave cellular network, e.g., $n_1=0$ and $n_2=0$. In that case, the user SINR outage probability is $1$, the probabilities of $n_1=0$ and $n_2=0$ are $\int_{R}^{\infty}f_{m}(r_m)dr_m=e^{-\lambda_m\pi R^2}$ and $\int_{r}^{\infty}f_{s}(r_s)dr_s=e^{-\lambda_s\pi r^2}$, respectively. Therefore, the all outage probability contain three cases which include no BSs in the two-tier network, the user associated with macro BS in the two-tier network and the user associated with pico BS in the two-tier network.
\begin{Lem}\label{prop:cell}
The all SINR outage probability of the two-tier heterogeneous mmWave cellular network is
\begin{align}
P_{\tout}(T)=P_{\tout,0}+P_{\tout,1}(T)+P_{\tout,2}(T),
\end{align}
where $P_{\tout,0}$ is the outage probability with no BS in the range, $P_{\tout,1}(T)$ is the outage probability with the user associated with a macro BS and $P_{\tout,2}(T)$ is outage probability with the user associated with a pico BS. They are given by as follows,
\begin{align}
P_{\tout,0}&=e^{-\lambda_m\pi R^2-\lambda_s\pi r^2},\\
P_{\tout,1}(T)&=\sum_{i=1}^{4}\int_{L_i^m}(1-e^{-\xi_m r_m^{\alpha_s}T}\mathcal{L}_{I_1}\mathcal{L}_{I_2})f_mf_s\,dr_sdr_m\nonumber\\
&+\sum_{j=1}^{2}e^{-\lambda_s\pi r^2}\int_{l_j^m}(1-e^{-\xi_m r_m^{\alpha_s}T}\mathcal{L}_{I_1})f_m\,dr_m,
\end{align}
\begin{align}
P_{\tout,2}(T)&=\sum_{i=1}^{4}\int_{L_i^s}(1-e^{-\xi_s r_s^{\alpha_s}T}\mathcal{L}_{I_3}\mathcal{L}_{I_4})f_mf_s\,dr_sdr_m\nonumber\\
&+\sum_{j=1}^{2}e^{-\lambda_m\pi R^2}\int_{l_j^s}(1-e^{-\xi_s r_s^{\alpha_s}T}\mathcal{L}_{I_4})f_s\,dr_s.
\end{align}

\end{Lem}

where $L_i^m$ and $L_i^s$ all denote the range of $r_m$ and $r_s$ that the user served by the macro BS and the pico BS. $l_j^m$ denotes the range of $r_m$ that the user served by the macro BS, $l_j^s$ denotes the range of $r_s$ that the user served by the pico BS.
$l_j^m$ and $l_j^s$ are given by
\begin{align}
l^m_1&=\{0\leq r_m\leq R_L\}, \quad l^m_2=\{R_L\leq r_m\leq R\}\\
l^s_1&=\{0\leq r_s\leq r_L\}, \quad l^s_2=\{r_L\leq r_s\leq r\}
\end{align}
$L_i^m$ and $L_i^s$ see Appendix \ref{app:range} for details.

\emph{Proof:} See Appendix \ref{app:cell} for details.\hfill$\square$

Note that it has been shown in \cite{stommWave}, \cite{gursoycoverage} that the mmWave systems are generally noise limited. For the noise limited systems, we can further simplify the cell SINR expressions.
\begin{Corr}
The outage probability of the user served by the macro BS and pico BS are given by
\begin{align}
\overline{P}_{out,s}^{m}(T) &= 1-e^{-\xi_m r_m^{\alpha_s}T},\\
\overline{P}_{out,s}^{s}(T) &=  1-e^{-\xi_s r_s^{\alpha_s}T}.
\end{align}
The all SNR outage probability is given by
\begin{align}
\overline{P}_{\tout}(T)=\overline{P}_{\tout,0}+\overline{P}_{\tout,1}(T)+\overline{P}_{\tout,2}(T),
\end{align}
where
\begin{align}
\overline{P}_{\tout,0}&=e^{-\lambda_m\pi R^2-\lambda_s\pi r^2},\\
\overline{P}_{\tout,1}(T)&=\sum_{i=1}^{4}\int_{L_i^m}(1-e^{-\xi_m r_m^{\alpha_s}T})f_mf_s\,dr_sdr_m\nonumber\\
&+\sum_{j=1}^{2}e^{-\lambda_s\pi r^2}\int_{l_j^m}(1-e^{-\xi_m r_m^{\alpha_s}T})f_m\,dr_m,\\
\overline{P}_{\tout,2}(T)&=\sum_{i=1}^{4}\int_{L_i^s}(1-e^{-\xi_s r_s^{\alpha_s}T})f_mf_s\,dr_sdr_m\nonumber\\
&+\sum_{j=1}^{2}e^{-\lambda_m\pi R^2}\int_{l_j^s}(1-e^{-\xi_s r_s^{\alpha_s}T})f_s\,dr_s.
\end{align}
\end{Corr}
\section{Numerical Results}
In this section, we evaluate the theoretical expressions numerically. The parameters are listed in Table I. We let $G=2$, e.g., the macro BS serves two groups users. $N_0(\mathrm{dBm})=-174+10\mathrm{log}_{10}(B)+\mathrm{NF(dB)}$, where $B$ and $\mathrm{NF}$ denote the bandwidth and noise figure, respectively.
\begin{table}
\caption{System Parameters}\label{tab:para}
\begin{center}
\begin{tabular}{ | c | c | c | }
\hline
parameter & definition & value \\
\hline
$M$ & Number of antennas & 128\\
\hline
 $K$&  number of users &10  \\
 \hline
$\theta_1$, $\theta_2$ & Each group AOA &  $-30^{\circ},0^{\circ}$ \\
\hline
$\Delta_1$, $\Delta_2$ & Each group AS &  $15^{\circ}, 15^{\circ}$ \\
 \hline
$f_c$ & carrier frequency & 28GHz\\
\hline
$P_m$ &The macro BS power  & $53$ dBm\\
 \hline
$P_s$ & The pico BS power &  $33$ dBm \\
 \hline
$B$ & Bandwidth &  1 GHz\\
\hline
$\mathrm{NF}$ &Noise figure & 10 dB\\
\hline
 $\alpha$ & Path loss & 4 \\
\hline
$R$, $r$ & tier 1 and tier 2 radiuses &  $200$ m ,$60$ m\\
\hline
$R_L,r_L$ & radius of LOS ball &  $20$ m, $20$ m\\
\hline
$\lambda_{m}, \lambda_{s}$ & densities of macro and pico BS & $10^{-5}, 10^{-4}$ \\
\hline
$\lambda_{u}$&  density of users &$10^{-3}$\\
\hline

\end{tabular}
\end{center}
\end{table}

\begin{figure}
    \centering
    \includegraphics[width=\figsize\textwidth]{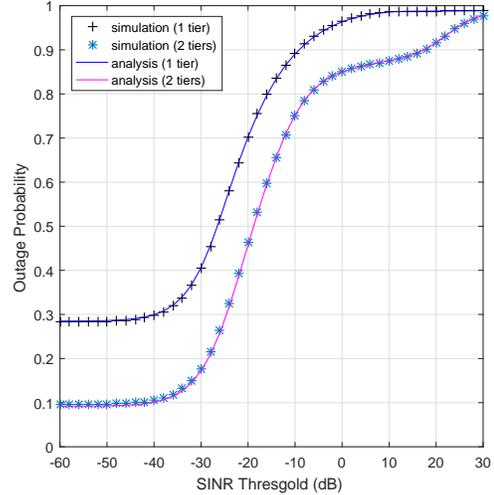}
    \caption{ The cell outage probability versus SINR threshold.}
    \label{fig:simulation-analysis}
\end{figure}

In Fig. \ref{fig:simulation-analysis}, we plot the all SINR outage probability with one-tier and two-tier networks. We can see that the simulation results match the analysis results perfectly whether it is the single tier or the two tiers, validating the theoretical analysis. Besides, compare the performance between the single tier and two tiers, employing two-tier heterogeneous network can improve the performance greatly than only employing macro BSs.

\begin{figure}
    \centering
    \includegraphics[width=\figsize\textwidth]{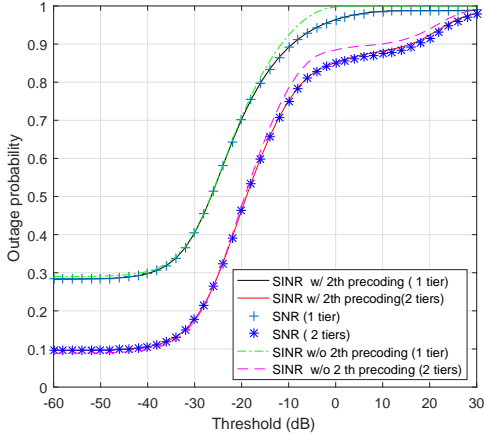}
    \caption{SNR and SINR comparison.}
    \label{fig:SNR-SINR}
\end{figure}

In Fig. \ref{fig:SNR-SINR}, we plot the all SINR and SNR outage probability with one-tier and two-tier networks. We first compare the SINR outage probability that with ZF precoding matrix as the second stage precoding with the SNR outage probability in single tier scenario and two tiers scenario. Note that the SINR and SNR outage probability in single tier and two tiers are almost same. Then we compare the SINR outage probability that with and without ZF precoding matrix as the second stage precoding in one-tier and two-tier networks, the system with second stage precoding have a better performance than the system without the second stage precoding.

\begin{figure}
    \centering
    \includegraphics[width=\figsize\textwidth]{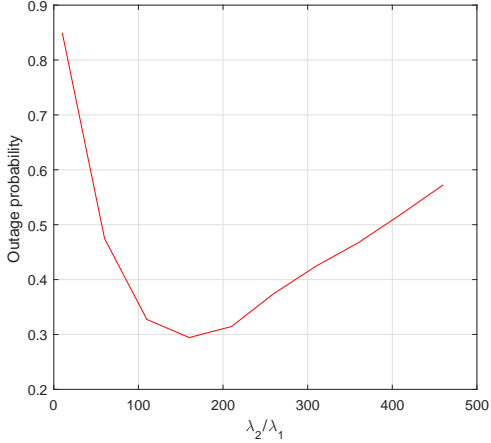}
    \caption{ The cell outage probability versus $\lambda_2/\lambda_1$ for SINR threshold of 0 dB.}
    \label{fig:density}
\end{figure}

In Fig. \ref{fig:density}, we plot the all SINR outage probability for two tiers network. From this figure, we can find the figure that the outage probability first decreases and then increases as $\lambda_2/\lambda_1$ increases. There is an optimal value for $\lambda_2/\lambda_1$ to achieve the smallest outage probability.

\section{Conclusion}
In this paper, we have provided a general analytical framework to compute the SINR outage probability of two-tier mmWave cellular networks employing JSDM. BSs of each tier and users are assume  to be distributed according to independent PPPs. We have analyzed the all SINR outage probability of the considered networks. Numerical evaluations have been provided to verify the theoretical analysis. We have shown that two-tier networks have better outage performance than single tier and the network using ZF as the second stage precoding outperforms the network without second stage precoding. Besides, we have found that the two-tier system is still noise-limited, which can help simplify the design of JSDM scheme.

\appendix
\subsection{Proof of Theorem \ref{prop:SINRoutage}}\label{app:SINRoutage}
According to \cite{interference}, $\Omega=||\C_{m,g}^{-1}\B_{m,g}^H||^2|\overline{\h}_{m,gk}^{H}\p_{m,gk}|^2$ follows Chi-squared distribution $\chi_2^2$. $\left|h_{\jmath}\right|^2$ is exponentially distributed unit mean. The outage probability is given by
\begin{align}
&\mathbb{P}(SINR_{mk}<T)\nonumber\\
&=\mathbb{P}\Bigg\{ \Omega<\frac{(I_1+I_2+N_0)r_{m}^{\alpha_s}T||\C_{m,g}^{-1}\B_{m,g}^H||^2}{P_m\kappa^2}\Bigg\}\nonumber\\
&=\mathbb{E}_{\{I_1,I_2\}}\Bigg\{\gamma\Big(1,\frac{(I_1+I_2+N_0)r_{m}^{\alpha_s}T||\C_{m,g}^{-1}\B_{m,g}^H||^2}{P_m\kappa^2}\Big)\Bigg\}\nonumber\\
&=\mathbb{E}_{\{I_1,I_2\}}\Bigg\{1-e^{-\frac{(I_1+I_2+N_0)r_{m}^{\alpha_s}T||\C_{m,g}^{-1}\B_{m,g}^H||^2}{P_m\kappa^2}}\Bigg\}\nonumber\\
&=1-e^{-\frac{N_0r_{m}^{\alpha_s}T||\C_{m,g}^{-1}\B_{m,g}^H||^2}{P_m\kappa^2}}\mathbb{E}_{\{I_1\}}\Big\{e^{-\frac{I_1r_{m}^{\alpha_s}T||\C_{m,g}^{-1}\B_{m,g}^H||^2}{P_m\kappa^2}}\Big\}\nonumber\\
&\times\mathbb{E}_{\{I_2\}}\Big\{e^{-\frac{I_2r_{m}^{\alpha_s}T||\C_{m,g}^{-1}\B_{m,g}^H||^2}{P_m\kappa^2}}\Big\}\nonumber\\
&=1-e^{-\frac{N_0r_{m}^{\alpha_s}T||\C_{m,g}^{-1}\B_{m,g}^H||^2}{P_m\kappa^2}}\mathcal{L}_{I_1}\mathcal{L}_{I_2},
\end{align}

where $\gamma(s,x)=\int_{0}^{x}t^{s-1}e^{-t}$. $\zeta_{\ell,g}=||\C_{\ell,g}^{-1}\B_{\ell,g}^H||^2|\overline{\h}_{\ell,gk}^{H}\p_{\ell,gk}|^2$ follows Chi-squared distribution $\chi_2^2$, $\mathcal{L}_{I_1}$ and $\mathcal{L}_{I_2}$ are given by
\begin{small}
\begin{align}
\mathcal{L}_{I_1}&=\mathbb{E}_{\{I_1,\Phi_1\}}\Big(e^{-\xi_m r_m^{\alpha_s}T\sum_{\ell\ne m}P_m\kappa^2 r_{\ell}^{-\alpha_s}\frac{\zeta_{\ell,g}}{||\C_{\ell,g}^{-1}\B_{\ell,g}^H||^2}}\Big)\nonumber\\
&=\mathbb{E}_{\{\Phi_1\}}\Big(\Pi_{\ell\ne m}\frac{1}{1+\xi_m r_m^{\alpha_s}TP_m\kappa^2\frac{r_{\ell}^{-\alpha_s}}{||\C_{\ell,g}^{-1}\B_{\ell,g}^H||^2}}\Big)\nonumber\\
&=\mathrm{exp}\Bigg(-\int_{r_m}^{a}\Big(1-\frac{1}{1+\xi_m r_m^{\alpha_s}TP_m\kappa^2x^{-\alpha_s}}\Big)\nonumber\\
&\times2\pi\lambda_m\alpha_sx^{\alpha_s}\,dx\Bigg)
\end{align}
\end{small}
\begin{small}
\begin{align}
\mathcal{L}_{I_2}&=\mathbb{E}_{\{I_2,\Phi_2\}}\Big( e^{-\xi_m r_m^{\alpha_s}T}\sum_{\jmath}P_s\kappa^2r_{\jmath}^{-\alpha_s}\left|h_{\jmath}\right|^2\Big)\nonumber\\
&=\mathbb{E}_{\{\Phi_2\}}\Big(\Pi_{\jmath}\frac{1}{1+\xi_m r_m^{\alpha_s}TP_s\kappa^2r_{\jmath}^{-\alpha_s}}\Big)\nonumber\\
&=\mathrm{exp}\Bigg(-\int_{r_m}^{a}\Big(1-\frac{1}{1+\xi_m r_m^{\alpha_s}TP_s\kappa^2x^{-\alpha_s}}\Big)\nonumber\\
&\times2\pi\lambda_s\alpha_sx^{\alpha_s}\,dx\Bigg)
\end{align}
\end{small}

Following the same reasoning, we can get $\mathcal{L}_{I_3}$ and $\mathcal{L}_{I_4}$.
\subsection{Proof of Theorem \ref{prop:cell}}\label{app:cell}
$P_{\tout,0}$ is equal to the probability that $n_1=0$ and $n_2=0$ at the same time.
The user is associated with the macro BS contain two cases. Case 1 is $n_2=0,n_1\neq 0$ and case 2 is $n_2\neq0,n_1\neq 0$ and $P_s\kappa^2r_{s}^{-\alpha_s}\leq P_m\kappa^2r_{m}^{-\alpha_s}$.

In that case the outage probability is
\begin{small}
\begin{align}
P_{\tout,11}&=P(n_2=0)\mathbb{E}_{r_m}(P_{out,s}^m)\nonumber\\
&=P(n_2=0)\Big(\mathbb{E}_{r_m}(P_{out,L}^m)+\mathbb{E}_{r_m}(P_{out,N}^m)\Big)\nonumber\\
&=\sum_{j=1}^{2}e^{-\lambda_s\pi r^2}\int_{l_j^m}(1-e^{-\xi_m r_m^{\alpha_s}T}\mathcal{L}_{I_1})f_m\,dr_m.
\end{align}
\end{small}
In this case, only macro BSs serve the user and the link between the macro BS and the user is either LOS link or NLOS link, $l_i$ denotes the range of LOS link or NLOS link.

In the case 2,$n_1=0$ and $n_2=0$ are equal to $0\leq r_s\leq r$ and $0\leq r_m\leq R$, respectively. The outage probability is
\begin{align}
P_{\tout,12}&=\mathbb{E}_{r_m}(P_{out,s}^m)\nonumber\\
&=\mathbb{E}_{r_m,r_s}\Big(P_{out,L}^m(r_m,r_s \mid P_s\kappa^2r_{s}^{-\alpha_L}\leq P_m\kappa^2r_{m}^{-\alpha_L})\Big)\nonumber\\
&+\mathbb{E}_{r_m,r_s}\Big(P_{out,L}^m(r_m,r_s \mid P_s\kappa^2r_{s}^{-\alpha_N}\leq P_m\kappa^2r_{m}^{-\alpha_L})\Big)\nonumber\\
&+\mathbb{E}_{r_m,r_s}\Big(P_{out,N}^m(r_m,r_s \mid P_s\kappa^2r_{s}^{-\alpha_L}\leq P_m\kappa^2r_{m}^{-\alpha_N})\Big)\nonumber\\
&+\mathbb{E}_{r_m,r_s}\Big(P_{out,N}^m(r_m,r_s \mid P_s\kappa^2r_{s}^{-\alpha_N}\leq P_m\kappa^2r_{m}^{-\alpha_N})\Big)
\end{align}
According the condition $P_s\kappa^2r_{s}^{-\alpha_s}\leq P_m\kappa^2r_{m}^{-\alpha_s}$, we can get the range of $r_m$ and $r_s$. The details see Appendix \ref{app:range}.
Therefore, the outage probability of the user served by the macro BS is
\begin{align}
P_{\tout,1}(x)=P_{\tout,11}(x)+P_{\tout,12}(x).
\end{align}

The outage probability of the user served by the pico BS is similar to the user served by the macro BS.

\subsection{Range for the association strategy} \label{app:range}
If $n1\neq0,n_2\neq0$ and the user is served by the macro BS, it must satisfy $P_s\kappa^2r_{s}^{-\alpha_s}\leq P_m\kappa^2r_{m}^{-\alpha_s}$. The link between the user and the macro BS or pico BS is either LOS link or NLOS link, e.g., it contain 4 cases that $P_s\kappa^2r_{s}^{-\alpha_L}\leq P_m\kappa^2r_{m}^{-\alpha_L}$, $P_s\kappa^2r_{s}^{-\alpha_N}\leq P_m\kappa^2r_{m}^{-\alpha_L}$, $P_s\kappa^2r_{s}^{-\alpha_L}\leq P_m\kappa^2r_{m}^{-\alpha_N}$ and $P_s\kappa^2r_{s}^{-\alpha_N}\leq P_m\kappa^2r_{m}^{-\alpha_N}$.

\begin{figure}
    \centering
    \includegraphics[width=\figsize\textwidth]{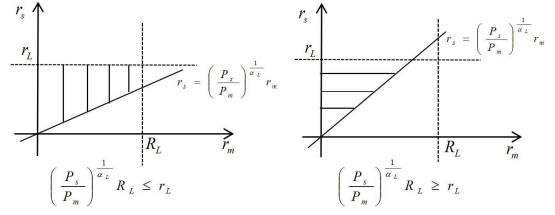}
    \caption{ The range of $P_s\kappa^2r_{s}^{-\alpha_L}\leq P_m\kappa^2r_{m}^{-\alpha_L}$.}
    \label{fig:tu1}
\end{figure}

\begin{figure}
    \centering
    \includegraphics[width=\figsize\textwidth]{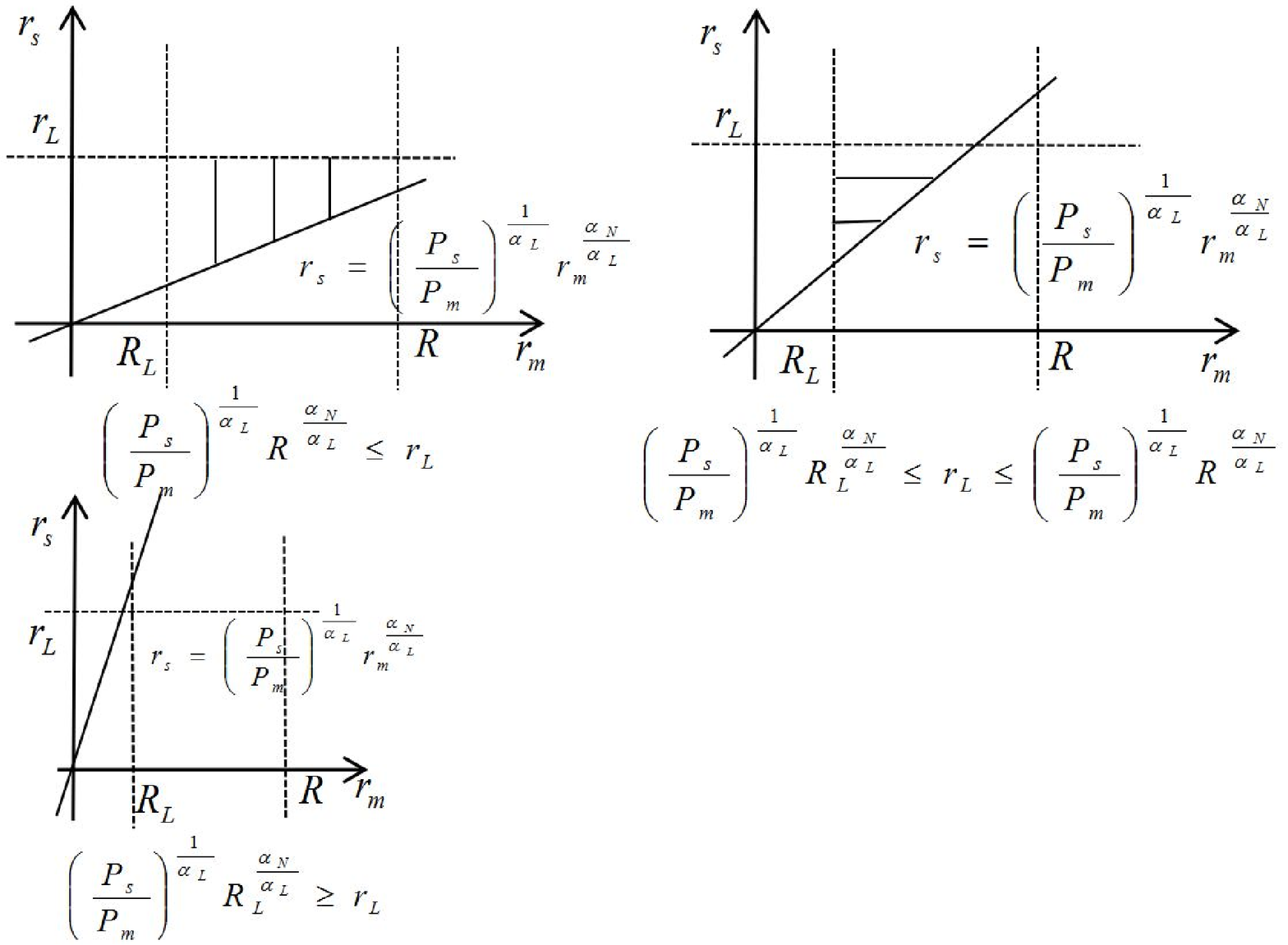}
    \caption{ The range of $P_s\kappa^2r_{s}^{-\alpha_N}\leq P_m\kappa^2r_{m}^{-\alpha_L}$.}
    \label{fig:tu2}
\end{figure}

\begin{figure}
    \centering
    \includegraphics[width=\figsize\textwidth]{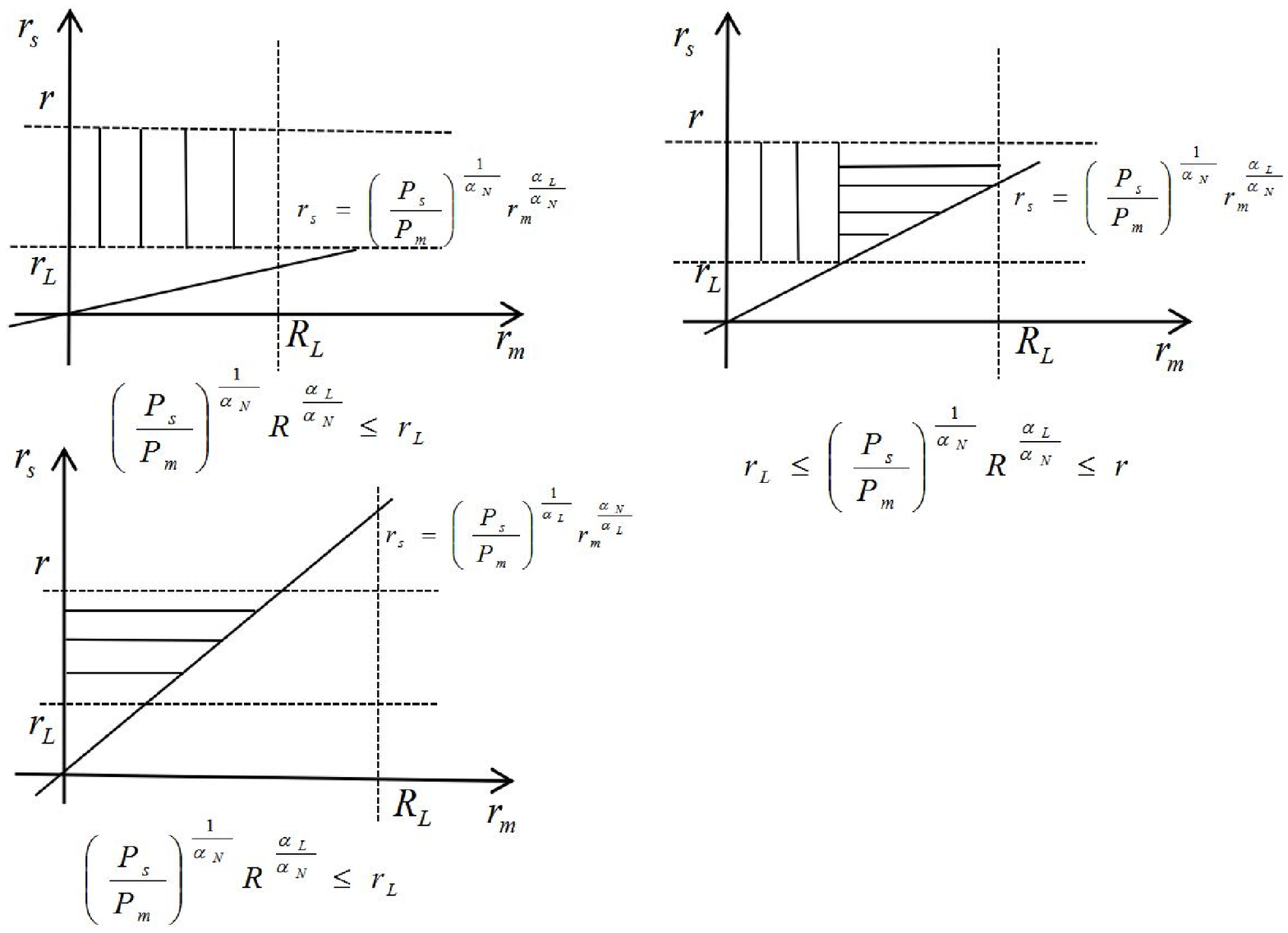}
    \caption{ The range of $P_s\kappa^2r_{s}^{-\alpha_L}\leq P_m\kappa^2r_{m}^{-\alpha_N}$.}
    \label{fig:tu3}
\end{figure}

\begin{figure}
    \centering
    \includegraphics[width=\figsize\textwidth]{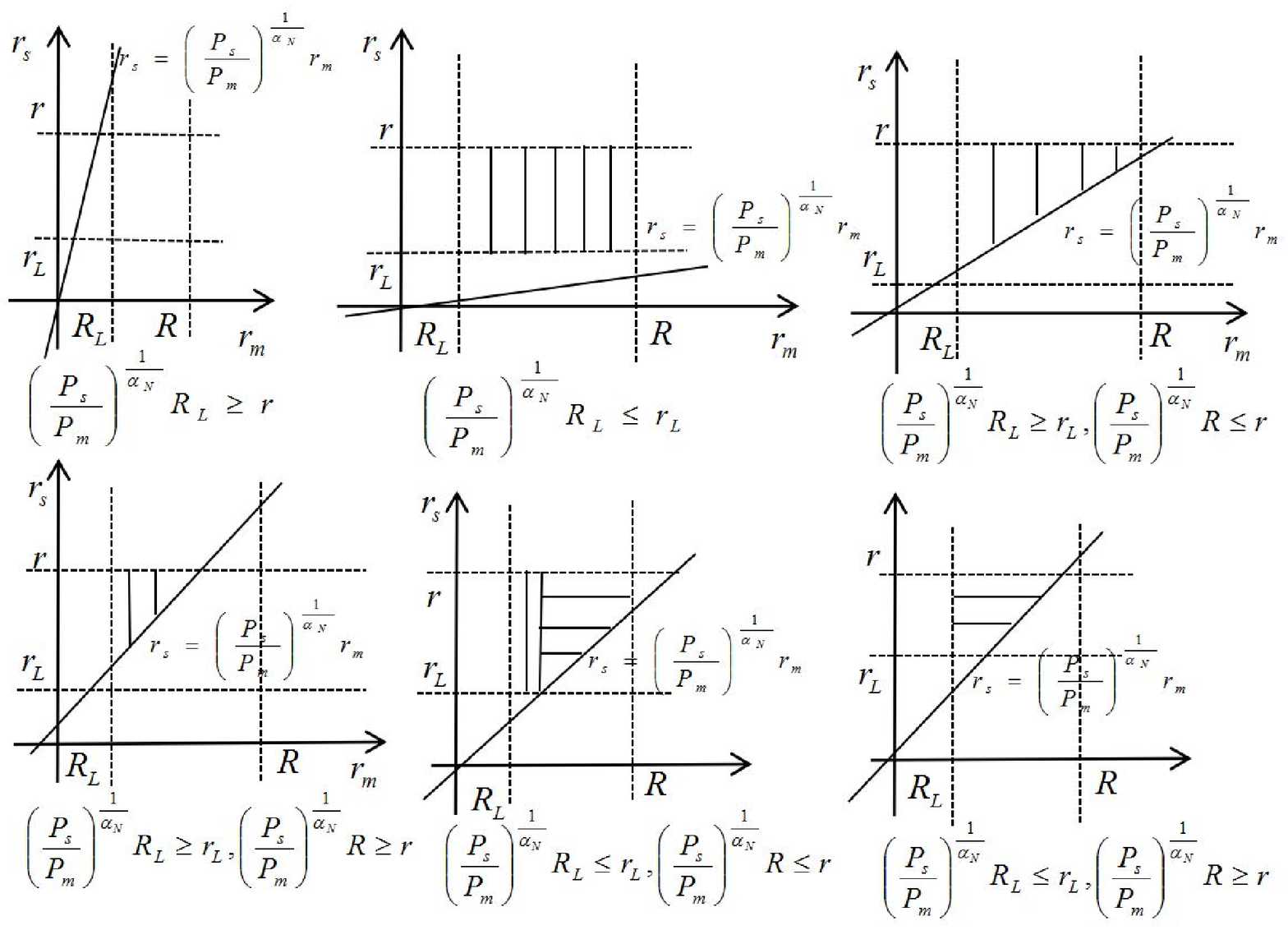}
    \caption{ The range of $P_s\kappa^2r_{s}^{-\alpha_N}\leq P_m\kappa^2r_{m}^{-\alpha_N}$.}
    \label{fig:tu4}
\end{figure}

For example, The range of $r_m$ and $r_s$ of condition $P_s\kappa^2r_{s}^{-\alpha_N}\leq P_m\kappa^2r_{m}^{-\alpha_N}$ is defined as
\begin{align}
L_4^m=\left\{
           \begin{aligned}
             &P_s\kappa^2r_{s}^{-\alpha_N}\leq P_m\kappa^2r_{m}^{-\alpha_N}\\
             &R_L\leq r_m\leq R\\
             &r_L\leq r_s\leq r
           \end{aligned}
      \right.
\end{align}
We use the knowledge of linear programming to find the feasible region. We compare different critical value to ensure interval value.
According to Fig. \ref{fig:tu1} - Fig. \ref{fig:tu4} that are the illustration of the $L^m_1 - L^m_4$, respectively, after some calculation, we can get the $L_1^m - L_4^m$. The feasible regions of $L^s_1 - L^s_4$ are the complementary set of $L^m_1 - L^m_4$, respectively.


\begin{small}
\begin{align}
L_1^m&=\left\{
           \begin{aligned}
           &\left\{
                \begin{aligned}
                    &(\frac{P_s}{P_m})^{\frac{1}{\alpha_L}}r_m \leq r_s \leq r_L\\
                    &0  \leq r_m \leq R_L
                \end{aligned}
                \right., &\mathrm{if} \;(\frac{P_s}{P_m})^{\frac{1}{\alpha_L}}R_L\leq r_L \\
                &\left\{
                \begin{aligned}
                    &0\leq r_s\leq r_L\\
                    &0  \leq r_m \leq (\frac{P_m}{P_s})^{\frac{1}{\alpha_L}}r_s
                \end{aligned}
                \right., &\mathrm{if} \;(\frac{P_s}{P_m})^{\frac{1}{\alpha_L}}R_L\geq r_L\\
           \end{aligned}
      \right.
\end{align}
\end{small}
\begin{small}
\begin{align}
L_2^m&=\left\{
           \begin{aligned}
           &\left\{
                \begin{aligned}
                    &(\frac{P_s}{P_m})^{\frac{1}{\alpha_L}}r_m^{\frac{\alpha_N}{\alpha_L}} \leq r_s \leq r_L\\
                    &R_L \leq r_m \leq R
                \end{aligned}
                \right., &\mathrm{if} \;(\frac{P_s}{P_m})^{\frac{1}{\alpha_L}}R^{\frac{\alpha_N}{\alpha_L}}\leq r_L \\
                &\left\{
                \begin{aligned}
                   &(\frac{P_s}{P_m})^{\frac{1}{\alpha_L}}r_m^{\frac{\alpha_N}{\alpha_L}} \leq r_s \leq r_L\\
                    &R_L \leq r_m \leq (\frac{P_m}{P_s})^{\frac{1}{\alpha_N}}r_L^{\frac{\alpha_L}{\alpha_N}}
                \end{aligned}
                \right., &\mathrm{if} \;\left\{
                                              \begin{aligned}
                                                     &(\frac{P_s}{P_m})^{\frac{1}{\alpha_L}}R^{\frac{\alpha_N}{\alpha_L}}\geq r_L\\
                                                     &(\frac{P_s}{P_m})^{\frac{1}{\alpha_L}}R_L^{\frac{\alpha_N}{\alpha_L}}\leq r_L
                                              \end{aligned}
                                        \right.\\
              &\varnothing, &\mathrm{if} \;(\frac{P_s}{P_m})^{\frac{1}{\alpha_L}}R_L^{\frac{\alpha_N}{\alpha_L}}\geq r_L
           \end{aligned}
\right.
\end{align}
\end{small}
\begin{small}
\begin{align}
L_3^m&=\left\{
           \begin{aligned}
           &\left\{
                \begin{aligned}
                   &r_L\leq r_s\leq r\\
                    &0  \leq r_m \leq R_L
                \end{aligned}
                \right., \mathrm{if} \;(\frac{P_s}{P_m})^{\frac{1}{\alpha_N}}R_L^{\frac{\alpha_L}{\alpha_N}}\leq r_L\\
                &\left\{
                \begin{aligned}
                        & \left\{
                              \begin{aligned}
                                      &r_L\leq r_s\leq r\\
                                      &0  \leq r_m \leq (\frac{P_m}{P_s})^{\frac{1}{\alpha_L}}r_L^{\frac{\alpha_N}{\alpha_L}}
                              \end{aligned}
                         \right.\\
                         &\left\{
                              \begin{aligned}
                                      & (\frac{P_m}{P_s})^{\frac{1}{\alpha_L}}r_L^{\frac{\alpha_N}{\alpha_L}}< r_m\leq R_L\\
                                      &(\frac{P_s}{P_m})^{\frac{1}{\alpha_N}}r_m^{\frac{\alpha_L}{\alpha_N}}\leq r_s\leq r
                              \end{aligned}
                         \right.
                \end{aligned}
                \right.,\mathrm{if}\;\left\{
                                              \begin{aligned}
                                                     &r_L\leq(\frac{P_s}{P_m})^{\frac{1}{\alpha_N}}R_L^{\frac{\alpha_L}{\alpha_N}}\\
                                                     &(\frac{P_s}{P_m})^{\frac{1}{\alpha_N}}R_L^{\frac{\alpha_L}{\alpha_N}}\leq r
                                              \end{aligned}
                                        \right.\\
                &\left\{
                \begin{aligned}
                       &0< r_m\leq (\frac{P_m}{P_s})^{\frac{1}{\alpha_L}}r_s^{\frac{\alpha_N}{\alpha_L}}\\
                       &r_L\leq r_s\leq r
                \end{aligned}
                \right.,\mathrm{if}\;(\frac{P_s}{P_m})^{\frac{1}{\alpha_N}}R_L^{\frac{\alpha_L}{\alpha_N}}\geq r\\
           \end{aligned}
\right.
\end{align}
\end{small}
\begin{small}
\begin{align}
L_4^m&=\left\{
           \begin{aligned}
              &\varnothing, &\mathrm{if} \;(\frac{P_s}{P_m})^{\frac{1}{\alpha_N}}R_L\geq r\\
              &\left\{
                \begin{aligned}
                   &r_L\leq r_s\leq r\\
                   &R_L \leq r_m \leq R
                \end{aligned}
                \right., &\mathrm{if} \;(\frac{P_s}{P_m})^{\frac{1}{\alpha_N}}R\leq r_L\\
                &\left\{
                \begin{aligned}
                   &(\frac{P_s}{P_m})^{\frac{1}{\alpha_N}}r_m\leq r_s\leq r\\
                   &R_L \leq r_m \leq R
                \end{aligned}
                \right., &\mathrm{if} \;\left\{
                                              \begin{aligned}
                                                     &(\frac{P_s}{P_m})^{\frac{1}{\alpha_N}}R_L\geq r_L\\
                                                     &(\frac{P_s}{P_m})^{\frac{1}{\alpha_N}}R\leq r
                                              \end{aligned}
                                        \right.\\
                &\left\{
                \begin{aligned}
                   &(\frac{P_s}{P_m})^{\frac{1}{\alpha_N}}r_m\leq r_s\leq r\\
                   &R_L \leq r_m \leq (\frac{P_m}{P_s})^{\frac{1}{\alpha_N}}r
                \end{aligned}
                \right., &\mathrm{if} \;\left\{
                                                \begin{aligned}
                                                       &(\frac{P_s}{P_m})^{\frac{1}{\alpha_N}}R_L\geq r_L\\
                                                       &(\frac{P_s}{P_m})^{\frac{1}{\alpha_N}}R\geq r
                                                \end{aligned}
                                        \right.\\
              &\left\{
                \begin{aligned}
                         &\left\{
                              \begin{aligned}
                                       &r_L\leq r_s\leq r\\
                                      &R_L\leq r_m\leq (\frac{P_m}{P_s})^{\frac{1}{\alpha_N}}r_L
                              \end{aligned}
                         \right.\\
                         &\left\{
                              \begin{aligned}
                                      &(\frac{P_s}{P_m})^{\frac{1}{\alpha_N}}r_m\leq r_s\leq r\\
                                      &(\frac{P_m}{P_s})^{\frac{1}{\alpha_N}}r_L\leq r_m\leq R
                              \end{aligned}
                         \right. \\
                \end{aligned}
                \right., &\mathrm{if} \;\left\{
                                                \begin{aligned}
                                                       &(\frac{P_s}{P_m})^{\frac{1}{\alpha_N}}R_L\leq r_L\\
                                                       &(\frac{P_s}{P_m})^{\frac{1}{\alpha_N}}R\leq r
                                                \end{aligned}
                                        \right.\\
                &\left\{
                \begin{aligned}
                   &r_L\leq r_s\leq r\\
                   &R_L \leq r_m \leq(\frac{P_m}{P_s})^{\frac{1}{\alpha_N}}r_s
                \end{aligned}
                \right., &\mathrm{if} \;\left\{
                                                \begin{aligned}
                                                       &(\frac{P_s}{P_m})^{\frac{1}{\alpha_N}}R_L\leq r_L\\
                                                       &(\frac{P_s}{P_m})^{\frac{1}{\alpha_N}}R\geq r
                                                \end{aligned}
                                        \right.\\
           \end{aligned}
\right.
\end{align}
\end{small}
\begin{small}
\begin{align}
L_1^s&=\left\{
           \begin{aligned}
           &\left\{
                \begin{aligned}
                    &0\leq r_s \leq (\frac{P_s}{P_m})^{\frac{1}{\alpha_L}}r_m \\
                    &0  \leq r_m \leq R_L
                \end{aligned}
                \right., &\mathrm{if} \;(\frac{P_s}{P_m})^{\frac{1}{\alpha_L}}R_L\leq r_L \\
                &\left\{
                \begin{aligned}
                    &0\leq r_s\leq r_L\\
                    &(\frac{P_m}{P_s})^{\frac{1}{\alpha_L}}r_s  \leq r_m \leq R_L
                \end{aligned}
                \right., &\mathrm{if} \;(\frac{P_s}{P_m})^{\frac{1}{\alpha_L}}R_L\geq r_L\\
           \end{aligned}
      \right.
\end{align}
\end{small}
\begin{small}
\begin{align}
L_2^s&=\left\{
           \begin{aligned}
           &\left\{
                \begin{aligned}
                    & 0 \leq r_s \leq (\frac{P_s}{P_m})^{\frac{1}{\alpha_L}}r_m^{\frac{\alpha_N}{\alpha_L}}\\
                    &R_L \leq r_m \leq R
                \end{aligned}
                \right., \mathrm{if} \;(\frac{P_s}{P_m})^{\frac{1}{\alpha_L}}R^{\frac{\alpha_N}{\alpha_L}}\leq r_L \\
            &\left\{
              \begin{aligned}
                    & 0 \leq r_s \leq r_L\\
                    &R_L \leq r_m \leq R
                \end{aligned}
                \right., \mathrm{if} \;(\frac{P_s}{P_m})^{\frac{1}{\alpha_L}}R_L^{\frac{\alpha_N}{\alpha_L}}\geq r_L\\
                &\left\{
                \begin{aligned}
                &\left\{
                              \begin{aligned}
                                      &0\leq r_s\leq r_L\\
                                      &(\frac{P_m}{P_s})^{\frac{1}{\alpha_N}}r_L^{\frac{\alpha_L}{\alpha_N}} \leq r_m\leq R
                              \end{aligned}
                         \right.\\
                         &\left\{
                              \begin{aligned}
                                      &R_L\leq r_m\leq (\frac{P_m}{P_s})^{\frac{1}{\alpha_N}}r_L^{\frac{\alpha_L}{\alpha_N}}\\
                                      &0\leq r_s\leq (\frac{P_s}{P_m})^{\frac{1}{\alpha_L}}r_m^{\frac{\alpha_N}{\alpha_L}}
                              \end{aligned}
                         \right. \\
                \end{aligned}
                \right., \mathrm{if} \;\left\{
                                                \begin{aligned}
                                                       &(\frac{P_s}{P_m})^{\frac{1}{\alpha_L}}R_L^{\frac{\alpha_N}{\alpha_L}}\leq r_L\\
                                                       &(\frac{P_s}{P_m})^{\frac{1}{\alpha_L}}R^{\frac{\alpha_N}{\alpha_L}}\geq r_L
                                                \end{aligned}
                                        \right.\\
           \end{aligned}
\right.
\end{align}
\end{small}
\begin{small}
\begin{align}
L_3^s&=\left\{
           \begin{aligned}
                &\varnothing, &\mathrm{if} \;(\frac{P_s}{P_m})^{\frac{1}{\alpha_N}}R_L^{\frac{\alpha_L}{\alpha_N}}\leq r_L\\
                &\left\{
                \begin{aligned}
                   & (\frac{P_m}{P_s})^{\frac{1}{\alpha_L}}r_L^{\frac{\alpha_N}{\alpha_L}}< r_m\leq R_L\\
                   &r_L\leq r_s\leq (\frac{P_s}{P_m})^{\frac{1}{\alpha_N}}r_m^{\frac{\alpha_L}{\alpha_N}}
                \end{aligned}
                \right.,&\mathrm{if}\;\left\{
                                              \begin{aligned}
                                                     &(\frac{P_s}{P_m})^{\frac{1}{\alpha_L}}R^{\frac{\alpha_N}{\alpha_L}}\geq r_L\\
                                                     &(\frac{P_s}{P_m})^{\frac{1}{\alpha_L}}R_L^{\frac{\alpha_N}{\alpha_L}}\leq r_L
                                              \end{aligned}
                                        \right.\\
                &\left\{
                \begin{aligned}
                   &(\frac{P_m}{P_s})^{\frac{1}{\alpha_L}}r_s^{\frac{\alpha_N}{\alpha_L}}< r_m\leq R_L\\
                   &r_L\leq r_s\leq r
                \end{aligned}
                \right.,&\mathrm{if}\;(\frac{P_s}{P_m})^{\frac{1}{\alpha_N}}R_L^{\frac{\alpha_L}{\alpha_N}}\geq r\\
           \end{aligned}
\right.
\end{align}
\end{small}
\begin{small}
\begin{align}
L_4^s&=\left\{
           \begin{aligned}
              &\left\{
                \begin{aligned}
                   &r_L\leq r_s\leq r\\
                   &R_L \leq r_m \leq R
                \end{aligned}
                \right., &\mathrm{if} \;(\frac{P_s}{P_m})^{\frac{1}{\alpha_N}}R_L\geq r\\
               &\varnothing, &\mathrm{if} \;(\frac{P_s}{P_m})^{\frac{1}{\alpha_N}}R\leq r_L\\
                &\left\{
                \begin{aligned}
                   &r_L\leq r_s\leq (\frac{P_s}{P_m})^{\frac{1}{\alpha_N}}r_m\\
                   &R_L \leq r_m \leq R
                \end{aligned}
                \right., &\mathrm{if} \;\left\{
                                              \begin{aligned}
                                                     &(\frac{P_s}{P_m})^{\frac{1}{\alpha_N}}R_L\geq r_L\\
                                                     &(\frac{P_s}{P_m})^{\frac{1}{\alpha_N}}R\leq r
                                              \end{aligned}
                                        \right.\\
                &\left\{
                \begin{aligned}
                &\left\{
                              \begin{aligned}
                                    &r_L\leq r_s\leq r\\
                                    &(\frac{P_m}{P_s})^{\frac{1}{\alpha_N}}r \leq r_m \leq R
                              \end{aligned}
                        \right.\\
                        &\left\{
                              \begin{aligned}
                                     &r_L\leq r_s\leq (\frac{P_s}{P_m})^{\frac{1}{\alpha_N}}r_m\\
                                     &R_L \leq r_m \leq (\frac{P_m}{P_s})^{\frac{1}{\alpha_N}}r
                              \end{aligned}
                         \right.\\
                \end{aligned}
                \right., &\mathrm{if} \;\left\{
                                                \begin{aligned}
                                                       &(\frac{P_s}{P_m})^{\frac{1}{\alpha_N}}R_L\geq r_L\\
                                                       &(\frac{P_s}{P_m})^{\frac{1}{\alpha_N}}R\geq r
                                                \end{aligned}
                                        \right.\\
              &\left\{
                \begin{aligned}
                   &r_L\leq r_s\leq(\frac{P_s}{P_m})^{\frac{1}{\alpha_N}}r_m\\
                   &(\frac{P_m}{P_s})^{\frac{1}{\alpha_N}}r_L\leq r_m \leq R
                \end{aligned}
                \right., &\mathrm{if} \;\left\{
                                                \begin{aligned}
                                                       &(\frac{P_s}{P_m})^{\frac{1}{\alpha_N}}R_L\leq r_L\\
                                                       &(\frac{P_s}{P_m})^{\frac{1}{\alpha_N}}R\leq r
                                                \end{aligned}
                                        \right.\\
                &\left\{
                \begin{aligned}
                       &r_L\leq r_s\leq r\\
                       &(\frac{P_m}{P_s})^{\frac{1}{\alpha_N}}r_s \leq r_m \leq R
                \end{aligned}
                \right., &\mathrm{if} \;\left\{
                                                \begin{aligned}
                                                       &(\frac{P_s}{P_m})^{\frac{1}{\alpha_N}}R_L\leq r_L\\
                                                       &(\frac{P_s}{P_m})^{\frac{1}{\alpha_N}}R\geq r
                                                \end{aligned}
                                        \right.\\
           \end{aligned}
\right.
\end{align}
\end{small}

\end{document}